# A Contextualist Decision Theory

## Saleh Afroogh

## I

**Contextualism vs. Inavriantism**

  Decision theorists propose a normative theory of rational choice. Traditionally, they assume that they should provide some *constant* and *invariant* principles as criteria for rational decisions, and indirectly, for agents. They seek a decision theory that invaribably works for all agents all the time. They believe that a rational agent should follow a certain principle, perhaps the principle of maximizing expected utility everywhere, all the time. As a result of the given context, these principles are considered, in this sense, *context-independent*.
  Furthermore, decision theorists usually assume that the relevant agents at work are ideal agents, and they believe that non-ideal agents should follow them so that their decisions qualify as rational. These principles are *universal* rules. I will refer to this *context-independent* and *universal* approach in traditional decision theory as **Invariantism**. This approach is, implicitly or explicitly, adopted by theories which are proposed on the basis of these two assumptions.
On the contrary, consider an alternative approach which doesn't assume that a decision theory is *context-independent* or *universal*. According to this new approach, which I call **contextulist decision theory,** the notion of rationality is relevant to context; and applies differently in variant contexts with distinct agents.

The important point is that, contextualist decision theory is not, directly, a theory about rational decision, neither is a rival theory for other decision theories; it can be best understood as a pragmatic metalinguistic theory about the usages of certain propositions involving terms such as "rationality", "rational decisions", "rational agents", etc. Put it in other words, contextualist decision theory doesn't propose any new theory regarding the nature or definition of "rationality" in decision theory. Instead, it assumes a notion of rationality or a theory of decision making; and then talks about the realization or truth conditions of that notion in different contexts.

Consider a smart agent who believes in the standard principle of *maximizing expected utility*, and thus, follows the standard formula of expected utility to make rational decisions in her life. However, her commitment to this formula and the accuracy of her calculation of the expected utility depends on pragmatic sensitivity in the context. Her being in a high or low-stake situations plays a central role in our evaluation, and in the legitimacy of ascribing rationality to her decision, and indirectly, to herself.

In the following section, I explain why *invariantism* is not a correct ground for decision theory; I shall propose some counterexamples to invariantism, and will show how we can explain them just in terms of the contextualist approach.

## II

### Counterexamples to invariantism:

I believe that there are several counterexamples to invariantism, and I think they can be best explained just in terms of contextualism.

**Counterexample (1):**
Consider these two lotteries;
1- Lottery A involves exactly two tickets. If ticket 1 is drawn you win 20 units of value, otherwise you lose 10. So, the expected utility of participating in this lottery would be:
   EU: ½ . 20 + ½ . (-10)= 10 – 5= 5

2- Lottery B also involves just two tickets. If you get ticket 1, then you will win 20,000,000 units of value, otherwise you lose 10,000,000. So, the expected utility of participating in this lottery would be:
   EU= ½ . 20,000,000 +1/2 . -10,000,000=5,000,000

According to the EU principle, it would be rational to prefer B to A; however, we intuitively believe that A is the only rational decision in this case. So, it is not always rational to follow the principle of maximizing expected utility principle. Sometimes we need to act against it. Therefore, it is completely relative to context.

It might be said that "utility" does not just consist in monetary values. The utility of an act would be the sum of all numbers assigned to the positive and negative values of the outcomes, in addition to the positive and negative monetary value. So, in lottery 2, we should consider the risk or regret which specifies some negative value as part of the utility of our action; and we should assign a relevant number, say for example (-30,000,000), to the risk as a negative value, and add it to (-10,000,000: negative monetary value if we lose), therefore, the EU of the second lottery would be much lower than the first one.

If we accept this interpretation of "utility", then this case would not constitute a counterexample to invariantism.

It seems that in order to provide a convincing counterexample, we need to introduce a case with a low expected utility, which is, nonetheless, rational relative to the context. In what follows, we will see that such an expectation is not realistic or required, and in fact such a counterexample is impossible.

The contextualist decision theorist doesn't claim that there can be contexts in which the EU of A is high than that of B, while it is intuitively more rational to do B. It seems that maximization of the utility - in its general sense, not just a technical formula - is an inseparable constituent of the notion of rationality, and we can say that *maximization of expected utility* for the notion of "rationality",in its instrumental sense, is like *factivity* for the notion of "knowledge". In other words, it would be contradictory to say:

- Decision "A" is instrumentally rational but it does not maximize the expected utility.

Or

- Decision A is more rational than B, but, decision B has a higher expected utility than A.

It is strongly intuitive that what is instrumentally rational for any agent should get the agent closer to her goal, and should satisfy at least some parts of her desires which form the goal.

Hence, this is not a correct counterexample. But, what does the contextualist decision theorist want to say? And what would be the proper counterexample to invariantism?

According to the contextualist decision theory, any acceptable decision theory, and the relevant notion of rationality (whatsoever) is relative to context; and pragmatic sensitivity and epistemic sensitivity are two determining factors in different contexts. We can explain this as follows.

First, we redefine all notions of "rationality" in terms of the three following constitutes, and say it is,

1- A feature of an act and, indirectly, of an agent;
2- Maximizing expected utility - in a general sense, not based on a technical reason or formula like Savage's - is a substantial part of that,
3- There is a reason (mainly formal) according to which we claim this act maximizes our expected utility or has more expected utility than the other.

And secondly, we divide these reasons (which is mentioned in 3) as follows:

a- The best or more technical reasons which can explain an ideal agent's preferences such as von Neumann and Morgenstern or Savag's theory, considering all possible state (more objective rules).
b- Some ordinary and internal reasons by non-ideal agents (more subjective rules).

The contexualist, along with the traditional invariantist, believes that a rational decision should maximize the expected utilities, however she says that "as the pragmatic stakes rise or the logical and epistemically precisions become more serious, the contextual standard gets more demanding." Like justification in epistemology which aims at providing a true belief, reason in decision theory aims at recognizing the closest decision to the right decision with the most expected utilities.

So a relevant counterexample to invariantism is not a case in which a rational decision maker decides to follow an action with less expected utility. In fact, that would be actually contradictory. A accurate and convincingcounterexample is a case in which an action is, according to a theory, say ordinary EUP, rational in one context and irrational in another context. Also an action is, according the simple calculation of EU (without considering all possible states), rational in one case and irrational in other cases. For more clarification, consider the following counterexamples.

**Example (2):**
Consider the non-ideal agent A: (by non-ideal agent I mean an agent who cannot represent her preferences without any contradiction such that can be explained by one of the above theories in a)
1- Consider person A who holds a lottery ticket with just two options (probability is ½). The two possible outcomes for A are + $10 and - $11). She participates in the lottery and her reason for this is just her subjective kinds of reason, and since stakes are low, it is not an irrational decision for A.
2- Consider person A who wants to choose the best option for her daughter's heart transplant surgery. She doesn't rely on her subjective reason (or a simple and naive decision theory) to decide in this high stake situation. She rightly believes that making decision, in this high-stake situation, without consulting with ideal agents would be irrational. In other words, due to the pragmatic sensitivity in this case, the rationality standards get more demanding and the only rational decision for that agent is consulting with some ideal agents and following them.

The important difference between this example and the first is that the agents in the first example are ideal agents who fully understand the principle of maximizing expected utility, and can correctly exhibit their preferences. In the second example the agents are non-ideal, however, that doesn't necessarily mean that they would be irrational in all of their decisions. It completely depends on different contexts.

**Example (3):**
Another still more commonsensical example is as follows.
- Consider A who is a local shopkeeper in College Station, and she wants to expand her business marketing. She decides to consult with a marketing counselor to get some advice. Making a commercial decision on the basis of advice by just one marketing counselor would be rational for a local shopping in College Station. However, such an approach would be irrational for owners of Google Incorporation who wants to develop their business. If they want to make some rational decisions for their business, they need to establish a more professional committee including some of the best decision theorists, economists and marketing counselors.

**III**

**The explanatory power of contextualism**

I explained how we can think of contextualism, as opposed to invariantism, as an alternative approach in decision theory. I also mentioned contextualism is not a rival for other decision theories. However, it might be said how it is possible to have an alternative approach which is not a rival for other decision theories. In fact, this problem traces back to different levels of explanation at work. On a more basic level, I take contextualism as opposed to invariantism (which is the common ground for all traditional decision theories). Hence, it is not a rival for decision theories. However, if we notice that contextualism puts some restrictions on all traditional decision theories, we can say that this new approach is contrary to all other theories, in their universal sense. It means that according to contextualism, we first reject the universality of other decision theories, and then we accept their applications just in some limited contexts. From this point of view, we can say that contextualism is a rival for all traditional decision theories, too.
Therefore, contextualism can explain much more commonsensical extensions of rational decisions in different contexts. The scope of application of each traditional theory is limited, although we can explain all commonsensical extensions in terms of contextualism, using anyone of traditional theories in their appropriate contexts. (Afroogh 2019)

I believe that contextualism can also explain some non-intuitive implications of invariantism. Some non-intuitive implications of invariantism are as follows, and we can avoid these implications if we adopt the contextualist approach.

Some non-intuitive implications of invariantism are as follows:

- Every decision theory which assumes invariantism would be so excessively narrow (i.e. explains just a limited and special group of commonsensical extensions) such that if you accept it, for example, say von Neumann and Morgenstern's theory, you should exclude not just all the people who cannot represent their preferences in terms of some special axioms, but still all other decision theorists who don't follow your principles. All of them will be considered as irrational agents in all their decisions making.
- As Paul Weirich proposed in *Models of decision makings*, some or most of these decision theories and principles are technical and non-commonsensical.
- Since every decision theory would propose a relevant notion of rationality, we face different notions of rationality without determining the relation between them. However, it seems that semantically there is just one meaning for rationality (in different levels) in the ordinary language.

Based on contextualism, howevr, we can say, without contradicting ourselves, that, Von Neumann and Morgenstern are rational people and, *at the same time* other decision theorists are also rational in their decision making. Moreover, we can also *legitimately* believe in both commonsensical and more technical decision theories and different levels of rationality. Furthermore, we can have different notions of rationalities in different contexts without any contradiction.

**High-standard and low-standard decision making**

I believe that the distinction between *ideal* and *non-ideal* agents is not realistic or precise. From a contextualist perspective, it is not the case that *ideal* agent would always follow the utility principles in all decision situations she faces. She does not always encounter high-stake decision situations which require high standards. Many of decision situations in everyday life are such that a decision making based on low standards and less precise principles would be enough to be rational in these cases. Consider an ideal agent and a decision theorist like Savage who should choose between two complicated decision theories and decide to follow the best one; and compare it with his approach when he should choose between two kinds of oranges to decide which one is the best to buy. As opposed to the first situation, in the second, he simply prefers to follow just the simple and more subjective maximization of expected utility principle without considering all the possible states and probabilities, which this is certainly a rational decision.

On the contrary, it is not the case that a non-ideal agent would always act irrationally based on her non-ideal decision principles. Sometimes she would face low-stake situations and it is enough for her to count as a rational agent *iff* she follows low-standards or subjective principles with less accurate results. Likewise, in a very sensitive situation, which requires high standards, she can decide rationally, because she is not intelligent enough to choose correctly, she can *suspend* any decision in such a situation and this *suspension* would be a rational decision for her in these situations, regardless of what the principles say.

### IV

### Conclusion

The main component of rationality is the commonsensical sense of maximizing the expected utility, not the technical formula, which is formulated in different decision theories and this is the common and shared section in all of these theories. Each of these theories explains just some extensions and according to the contextualist approach, which is a pragmatic one and is not a rival for other theories, we can explain all commonsensical extensions of different usages of the terms related to rationality - such as rational agent, rational action, rational idea, etc.
The last but not the least point is related to an important question. Why does an agent forexample, say an ideal agent, in different contexts, and based on pragmatic or epistemic sensitivity, invoke a certain decision theory? Is it arbitrary or not? It seems to me that even selecting an appropriate theory in a certain decision situation, in different contexts, could be explained in terms of contextualism and the commonsensical notion of maximizing expected utility.
Recall the shopkeeper in example 3 who is going to pay money to for meeting with just one marketing adviser to improve her business. It is rational for her to rely on the adviser. However, it is not rational for a big company to accept her business. A big company needs to hire more professional marketing experts to find some more advanced decision theory and more exact analysis considering almost all possible conditions. How can we explain the difference at work? It seems that both agents - the local shopkeeper and the Google Incorporation - first, calculate the simple maximizing expected utility to determine which strategy (consulting with just one adviser or hiring more professional counselors) would be the best one regarding their contexts. For a local shopkeeper, it will be too costly and time-consuming to follow more exact decision theories with more exact analyses. Therefore, it is not rational for her to do this. However, regarding the considerable incomes of the Google Incorporation, it would be fully rational to pay a lot of money to find the best decision theory and make the best decisions for their business.

**Reference:**

Patrick, Rysiew (2016) Epistemic Contextualism in, *The Stanford Encyclopedia of Philosophy.*

Weirich, Paul. (2015), *Models of decision makings,* Cambridge University Press.

Afroogh, Saleh. Contextual Reason and Rationality. Diss. 2019

Afroogh, Saleh (2019). Contextual Reason and Rationality. Master's thesis, Texas A&M University. Available electronically from http : / /hdl .handle .net /1969 .1 /186349.

10.13140/RG.2.2.21462.06726